# Similarity Measures, Author Cocitation Analysis, and Information Theory

Loet Leydesdorff

Science & Technology Dynamics, University of Amsterdam

Amsterdam School of Communications Research (ASCoR)

Kloveniersburgwal 48, 1012 CX  Amsterdam, The Netherlands

loet@leydesdorff.net ; http://www.leydesdorff.net

**Abstract**

The use of Pearson's correlation coefficient in Author Cocitation Analysis was compared with Salton's cosine measure in a number of recent contributions. Unlike the Pearson correlation, the cosine is insensitive to the number of zeros. However, one has the option of applying a logarithmic transformation in correlation analysis. Information calculus is based on both the logarithmic transformation and provides a non-parametric statistics. Using this methodology one can cluster a document set in a precise way and express the differences in terms of bits of information. The algorithm is explained and used on the data set which was made the subject of this discussion.

**Introduction**

In a provocative study, Ahlgren *et al*. (2003) questioned the use of Pearson's Correlation Coefficient as a similarity measure in author cocitation analysis (ACA) with the argument that this measure is sensitive for zeros. Analytically, the addition of zeros to two variables should add to their similarity, but the authors show with empirical examples that this addition can depress the correlation coefficient between these variables. Salton's cosine is suggested as a possible alternative because this similarity measure is insensitive to the addition of zeros (Salton & McGill, 1983).

In a reaction White (2003) defended the use of the Pearson correlation hitherto in ACA with the pragmatic argument that the differences between using different similarity measures can be neglected in the research practice. He illustrated this with dendrograms and mappings using Ahlgren *et al.*'s own data. Bensman (2004) contributed to the discussion with a letter in which he argued for using Pearson's *r* for additional reasons. Unlike the cosine, Pearson's *r* is embedded in multivariate statistics and because of the normalization implied this measure allows for negative values. The problem with the zeros can be solved by applying a logarithmic transformation to the data. In his opinion, this transformation is anyhow advisable in the case of a bivariate normal distribution.

Leydesdorff & Zaal (1988) experimented with comparing results of using various similarity criteria—among which the cosine and the correlation coefficient—and different clustering algorithms for the mapping. Indeed, the differences between using the Pearson's *r* or the cosine were also minimal in our case. However, our study was mainly triggered by concern about the use of single linkage clustering in the ISI's *World Atlas of Science* (Small & Sweeney, 1985; Small *et al.*, 1985). The choice for this algorithm had been made by the ISI for technical reasons given the computational limitations of that time.

Single linkage clustering is well-known for a tendency to link areas of high density together to one supercluster because of accidental in-between points (Everitt, 1974). I argued that the cocitation clusters in the *Atlas of Science* were sometimes confounded by this so-called effect of 'chaining.' For example, when Small *et al*. (1985) claimed that the larger part of the natural sciences is 'interdisciplinary' with chemistry 'to be considered the model of an interdisciplinary science,' this result could be considered as an effect of using the wrong algorithm (Leydesdorff, 1987).

The differences between using Pearson's Correlation Coefficient and Salton's cosine are marginal in practice because the correlation measure can also be considered as a cosine between normalized vectors (Jones & Furnas, 1987). The normalization is sensitive to the zeros, but as noted this can be repaired by the logarithmic transformation. More generally, however, it remains most worrisome that one has such a wealth of both similarity criteria (e.g., Euclidean distances, the Jaccard index, etc.) and clustering algorithms (e.g., single linkage, average linkage, Ward's mode, etc.) available that one is able to generate almost any representation from a set of data (Oberski, 1988).

The problem of how to estimate the number of clusters, factors, groups, dimensions, etc. is a pervasive one in multivariate analysis. If there are no *a priori* theoretical reasons—as is usually the case in exploratory uses of these techniques—such decisions tend to remain somewhat arbitrary. In factor analysis, methods such as visual inspection of the scree plot or a cut-off at certain eigenvalues are common practice. In cluster analysis and multi-dimensional scaling, decisions based upon visual inspection of the results are common.

Small & Sweeney (1985), for example, have proposed 'variable level clustering,' that is, in essence the adaptation of the clustering level to the density of the cluster involved; the search for a formal criterion is thus replaced by a procedural one. This practice was later implemented in the French system for co-word clustering LEXIMAPPE (Callon *et al*., 1986; Courtial, 1989), but the results of this system could not be validated when using Shannon's (1948) information theory (Leydesdorff, 1992).

## 2. An information theoretical approach

Can an exact solution be provided for the problem of the decomposition? I submit that information theory can be elaborated into statistical decomposition analysis (Theil, 1972) and that this methodology provides us with clear criteria for the dividedness (Leydesdorff, 1991, 1995). Dividedness and aggregation can both be expressed in terms of bits of information. I shall show that a function for the dividedness can then be maximized.

In general, disaggregation of a set in *g* groups can be described with the following formula:

$$H = H_0 + \Sigma_g P_g H_g \qquad (1)$$

in which H is the expected information content (probabilistic entropy) of the aggregated distribution, and $P_g$ the probability of each of the groups which as a subset has an uncertainty equal to the respective $H_g$s. The 'in between group entropy' $H_0$ is a measure of the specificity that prevails at the level of the subsets, and thus it should be possible to use it as a measure for the quality of clustering.

The right-hand term of the above equation ($\Sigma_g P_g H_g$) is equal to the entropy of a variable (*n*) under the condition of a grouping variable (*m*): *H(n/m)*. The left-hand term of Equation 1, $H_0$, is therefore, equal to *H(n) - H(n/m)*, which is the uncertainty in *n* that is *not* attributable to the uncertainty within the groups, or in other words the *transmission* (*mutual information*) of the grouping variable *m* to *n* (that is to be grouped). The larger this transmission, the more reduction of uncertainty there will be among the groups, and therefore the better the groups will be in terms of the homogeneity of their distributions. However, by definition:

$$H(n|m) = H(n,m) - H(m) \qquad (2)$$

Since $H_0 = H(n) - H(n|m)$ (see above), this implies:

$$H_0 = H(n) + H(m) - H(n,m) \qquad (3)$$

In other words, the increase of $H_0$ if one distinguishes an additional group (cluster, factor, etc.) is composed of a part that is dependent only on the grouping variable (*H(m)*), and a part which is dependent on the interaction between the grouping variable *m* and the grouped variable *n*. The interaction between the two variables makes *H(n,m)* smaller than the sum of *H(n)* and *H(m)*. Given a number of variables *n* to be grouped, the question thus becomes: for which value of *m* does the function *{H(m) - H(n,m)}*, and consequently $H_0$ as an indicator of the dividedness, reach a maximum? This problem can be solved numerically by recursive reallocation of the cases into all possible groupings. The (normalized) maximization of the $H_0$ thus provides an unambiguous criterion for the quality of the attribution. *Q.e.d.*

Let me formulate the argument also more intuitively: If we divide one group into two subgroups *i* and *j*, using $H_{ij} = H_0 + P_i H_i + P_j H_j$, the aggregated $H_{ij}$ may be larger than both $H_i$ and $H_j$, or larger than one of them and smaller than the other. (The two groups cannot be both larger than $H_{ij}$, since the 'in-between group uncertainty' $H_0$ is necessarily larger than or equal to zero.) The case of $H_i < H_{ij} < H_j$ corresponds to the removal of the more than average heterogeneous case(s) into a separate subgroup: therefore, this new subgroup has a higher uncertainty, and the remaining subgroup becomes more homogeneous than the original group. This is always possible, but it is not clustering.

Clustering entails by definition the notion of reducing uncertainty in *both* subgroups. Therefore, we may define 'divisive clustering' as the case where both new subgroups have a lower expected information content than the undivided group. Note that the above justification of the division is based only on the right-hand term of the formula for disaggregation ($\Sigma_g P_g H_g$ in Equation 1). The value of the left-hand term ($H_0$), however, is sensitive both to the number of groups—since each further division adds to $H_0$ unless the two groups have similar $H_g$s—and to the quality of the attribution of cases to groups given a certain number of groups. These two questions—(1) concerning the number of groups, and (2) concerning the attribution of cases to groups—can be studied independently, given the two terms in Equation 3.

The possible number of attributions of *n* variables to *m* groups (*m* < *n*) increases so rapidly with the number of units and the number of groups that systematic comparison of all possible combinations can imply heavy computation. In practice, this type of repetitive approach to the data, which is characteristic of information theory (Krippendorff, 1986), can be programmed in DO WHILE-loops. Note that this approach takes the data as given and not as a result of an underlying process. Thus, there is no assumption required about the shape of the distribution.

First, we investigate whether the setting apart of any of the cases leads to two subgroups, both of which have lower $H_g$s than the overall H. If so, we begin with the one which leads to the highest $H_0$, and systematically evaluate whether the addition of other cases to this one subgroup leads to a further increase of $H_0$, etc. Once we have investigated all the possibilities and decided upon the best division into two subgroups, the analysis can be repeated for the two subgroups respectively.

After normalization of $H_0$ in terms of the grand sum of the matrix, a dendrogram can be constructed, which is exact both in terms of the vertical distances between the nodes and in terms of where to draw the line above which further division leads to subgroups that are not both lower in their entropy than their respective aggregates. This level in the graph corresponds to a maximum for $H_0$.

**3. Application to the ACA dataset of Ahlgren *et al*. (2003)**

Like White (2003), I applied the above algorithm to the dataset which was published by Ahlgren *et al*. (2003) in Table 7, at p. 555. Since the algorithm proposed above is not included in standard software packages I had to construct a visualization in Excel based on the numeric output. Figure 1 provides the results of the analysis. These results are in some important respects different from the results presented by White (2003). For example, 'Nederhof' and 'Price' were never placed in the same leave of the dendrogram in White's results, but these two variables are among the highest correlated ones in the matrix with a Pearson's $r = 0.837$ and the cosine $= 0.904$.[1]

---

[1] Ahlgren et al. (2003, at p. 556) provide for this Pearson correlation the value of 0.86. However, this higher value is based on treating the diagonal values as missing data (ibid., p. 554). Because the diagonal values are outlyers, this value is overestimated. The approach of including the diagonal value (in this study) can be considered as conservative because the Pearson correlation is then underestimated. Normalizations for the diagonal values have been proposed by Price (1981) and Noma (1982).

From a substantive perspective, one may note as remarkable that Schubert and Van Raan are placed in the same leaf of the dendrogram provided in Figure 1, while Nederhof is placed with Braun and Schubert in another. This result is counterintuitive. However, it reveals the structure in the data more than a representation which is selected because it corresponds with our a priori knowledge of the field. The three-dimensional factor solution of the scientometric subgroup in this data (Table 1), for example, reveals this same structure.

**Rotated Component Matrix**

|  | Component | | |
| --- | --- | --- | --- |
|  | 1 | 2 | 3 |
| *Braun* | **.866** | .129 | .226 |
| *Glanzel* | **.819** | .226 | .253 |
| *Nederhof* | **.751** | .400 | .389 |
| *Moed* | .679 | .438 | .357 |
| *Narin* | .668 | .373 | .298 |
| *Price* | .657 | .541 | .243 |
| *Cronin* | .200 | **.900** | .264 |
| *Callon* | .333 | **.858** | .246 |
| *Tijssen* | .535 | .679 | .306 |
| *Schubert* | .328 | .130 | **.855** |
| *VanRaan* | .279 | .378 | **.761** |
| *Leydesdorff* | .317 | .471 | **.647** |

**Table 1**: Three factor solution for the scientometric subgroup in the data of Ahlgren *et al*. (2003). Extraction Method: Principal Component Analysis. Rotation Method: Varimax with Kaiser Normalization. Rotation converged in 6 iterations.

The information-theoretical approach does not only take the non-zero values into account like the cosine, but it also includes the information about the redundancy provided by the zeros in the matrix. This structural information is required for the multivariate analysis as Bensman (2004) noted.[2] However, Ahlgren *et al*.'s (2003) problem with the zeros affecting the Pearson correlation—although one needs this measure for the parametric approach—can also not be denied. Information theory enables us to combine the non-parametrical approach with addressing the information contained in the zeros.

---

[2] The zeros provide information about the hyper-planes that span the system and should therefore not be discarded as non-information (Mokken, *personal communication*).

The major advantage of using information theory for the clustering is the collapsing of two parameter spaces, notably for the similarity criteria and the clustering algorithms. One does not need another decision about the clustering algorithm (in addition to the decision about the similarity criterion) because information theory provides a calculus. The solution is unique (since exact) using a single criterion, notably the maximization of the in-between group uncertainty. All grouping and dividedness can be expressed precisely in terms of the bits of information contained in the data. The results of the precise (that is, algorithmic) decomposition can perhaps even be used as a yardstick for the evaluation of geometrical visualizations and other (e.g., parametric) estimations. Unfortunately, the measure is not yet available in standard software packages like SPSS, and therefore not so easily available as more standard routines.

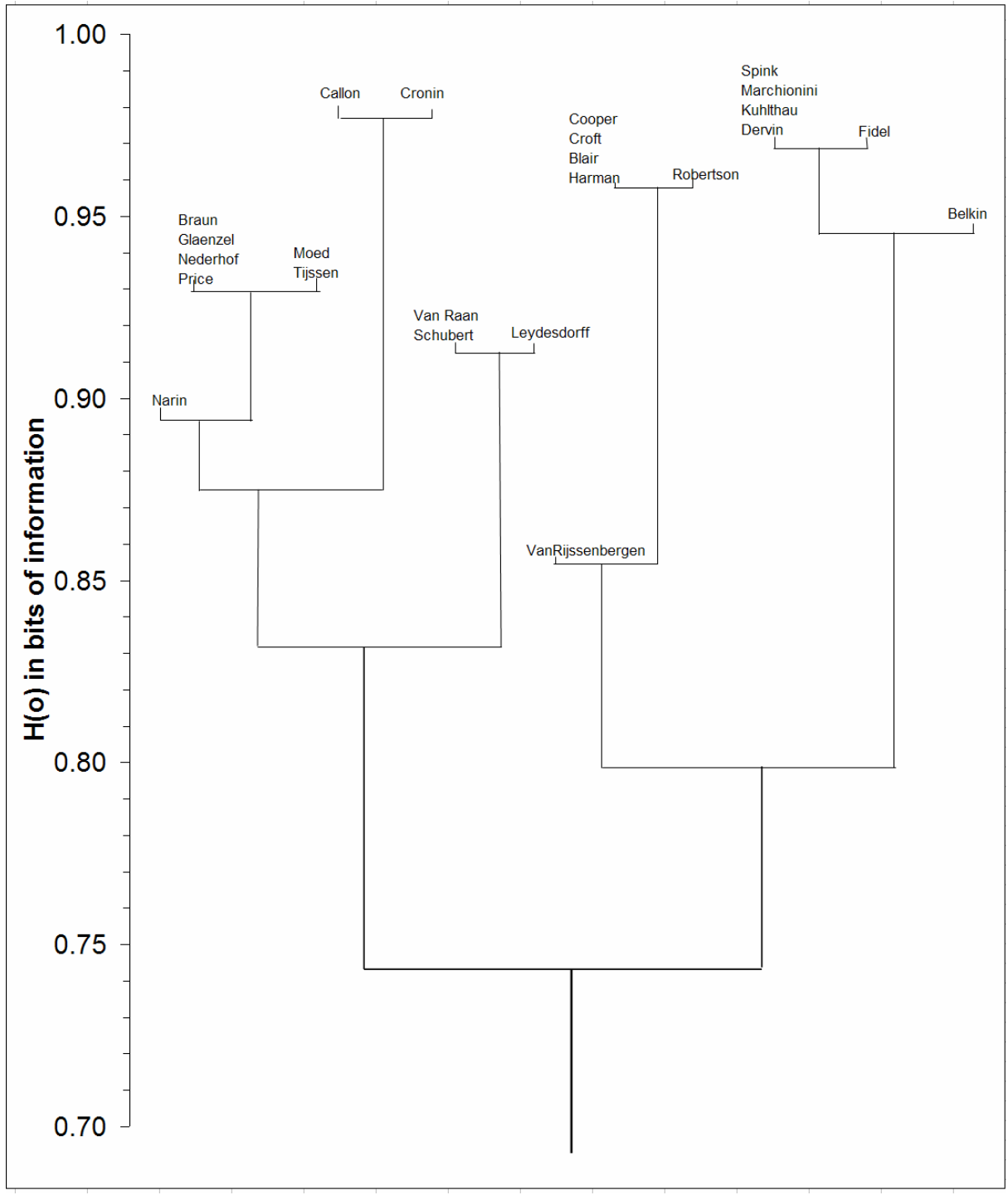

**Figure 1**

Divisive clustering of author co-citation matrix using information theory